\documentclass[12 pt,twoside,aps,prd,amsmath,amssymb,
tightenlines,showpacs,showkeys,eqsecnum]{revtex4-1}
\usepackage{mathptmx}
\usepackage{bm}
\usepackage{graphicx}

\usepackage{hyperref}
\newcommand {\G}{\Gamma}

\def \myfigures #1#2#3#4#5#6#7#8
{\begin{figure}[ht]
    \begin{center}
        \includegraphics[width=#2 \textwidth]{#1.eps}
        \hfill
        \includegraphics[width=#6 \textwidth]{#5.eps}
        \parbox[t]{#4\textwidth}{\caption {#3}\label{#1}}
        \hfill
        \parbox[t]{#8\textwidth}{\caption {#7}\label{#5}}
    \end{center}
\end{figure} }

\def\myfigure #1#2#3#4
{\begin{figure}[ht]\begin{center}
\includegraphics[width=#2 \textwidth]{#1.jpg}
\parbox[t]{#4\textwidth}{\caption{#3}\label{#1}}
\end{center}\end{figure}}

\begin{document}
\baselineskip -24pt
\title{Bianchi type-V dark energy model with varying EoS parameter}
\author{Bijan Saha}
\affiliation{Laboratory of Information Technologies\\
Joint Institute for Nuclear Research, Dubna  \\ 141980 Dubna, Moscow
region, Russia} \email{bijan@jinr.ru}
\homepage{http://bijansaha.narod.ru}

\begin{abstract}

Within the scope of an anisotropic Bianchi type-V cosmological model
we have studied the evolution of the universe. The assumption of a
diagonal energy-momentum tensor leads to some severe restriction on
the metric functions, which on its part imposes restriction on the
components of the energy momentum tensor. This model allows
anisotropic matter distribution. Further using the proportionality
condition that relates the shear scalar $(\sigma)$ in the model is
proportional to expansion scalar $(\vartheta)$ and the variation law
of Hubble parameter, connecting Hubble parameter with volume scale.
Exact solution to the corresponding equations are obtained. The EoS
parameter for dark energy as well as deceleration parameter is found
to be the time varying functions. A qualitative picture of the
evolution of the universe corresponding to different of its stages
is given using the latest observational data.
\end{abstract}



\maketitle

\bigskip

\section{Introduction}

The discovery of late time accelerating mode of expansion of the
Universe in one hand gave a boost to observational cosmology, at the
same time posing new challenges to cosmologists. Since its discovery
a number of models are offered to explain this phenomenon. Most of
the dark energy models such as quintessence, Chaplygin gas etc. are
simulated in analogy with the cosmological constant that gives rise
to a negative pressure. In doing so a constant EoS parameter was
considered. Recently in a number of papers different cosmological
models with time dependent EoS parameter was studied
\cite{APSAPSSbi0,APSCPL,PASIJTPbi,PASAPSS,SAPAPSS,YSAPSSbi,ECCADE-BVI}.
The aim of the current paper is to extend that study for a Bianchi
type-V cosmological model. It should be noted that a BV model can be
deduced from a BVI with some suitable choice of spatial dependence
of the metric function. A Bianchi type-V model describes an
anisotropic but homogeneous Universe. This model was studied by
several authors \cite{BVI,GCBVI2010,svBVI,Hoogen,Socorro,Weaver1},
specially due to the existence of magnetic fields in galaxies which
was proved by a number of astrophysical observations. Whereas, some
dark energy model within the scope of a BV cosmology was studied in
\cite{Yadav2011}.

\section{Basic equations}

Bianchi type-V model given be given by \cite{BVI,GCBVI2010}
\begin{equation}
ds^2 = dt^2 - a_1^{2} e^{-2mz}\,dx^{2} - a_2^{2} e^{-2mz}\,dy^{2} -
a_3^{2}\,dz^2, \label{bv}
\end{equation}
with $a_1,\,a_2,\,a_3$ being the functions of time only. Here $m$ is
some arbitrary constants and the velocity of light is taken to be
unity. Here we consider the case when the energy momentum tensor has
only non-trivial diagonal elements, i.e.
\begin{eqnarray}
T_\alpha^\beta = {\rm diag}[T_0^0,\,T_1^1,\,T_2^2,\,T_3^3]
\label{emt}
\end{eqnarray}

Einstein field equations for the metric \eqref{bv} on account of
\eqref{emt} have the form \cite{BVI}

\begin{subequations}
\label{einbv}
\begin{eqnarray}
\frac{\ddot a_2}{a_2} +\frac{\ddot a_3}{a_3} +\frac{\dot
a_2}{a_2}\frac{\dot
a_3}{a_3} - \frac{m^2}{a_3^2} &=& \kappa T_{1}^{1}, \label{11bv}\\
\frac{\ddot a_3}{a_3} +\frac{\ddot a_1}{a_1} +\frac{\dot
a_3}{a_3}\frac{\dot
a_1}{a_1} - \frac{m^2}{a_3^2} &=& \kappa T_{2}^{2}, \label{22bv} \\
\frac{\ddot a_1}{a_1} +\frac{\ddot a_2}{a_2} +\frac{\dot
a_1}{a_1}\frac{\dot
a_2}{a_2} - \frac{m^2}{a_3^2} &=& \kappa T_{3}^{3}, \label{33bv}\\
\frac{\dot a_1}{a_1}\frac{\dot a_2}{a_2} +\frac{\dot a_2}{a_2}
\frac{\dot a_3}{a_3} + \frac{\dot a_3}{a_3}\frac{\dot a_1}{a_1} -
3\frac{m^2}{a_3^2} &=&
\kappa T_{0}^{0}, \label{00bv}\\
\frac{\dot a_1}{a_1} + \frac{\dot a_2}{a_2} - 2 \frac{\dot a_3}{a_3}
&=& 0. \label{03bv}
\end{eqnarray}
\end{subequations}

We define the spatial volume of the model \eqref{bv} as
\begin{equation}
V = a_1 a_2 a_3, \label{Vdef}
\end{equation}
and the average scale factor as
\begin{equation}
a = V^{1/3} = (a_1 a_2 a_3)^{1/3}. \label{adef}
\end{equation}

Let us now find expansion and shear for BVI metric. The expansion is
given by
\begin{equation}
\vartheta = u^\mu_{;\mu} = u^\mu_{\mu} + \G^\mu_{\mu\alpha}
u^\alpha, \label{expansion}
\end{equation}
and the shear is given by
\begin{equation}
\sigma^2 = \frac{1}{2} \sigma_{\mu\nu} \sigma^{\mu\nu},
\label{shear}
\end{equation}
with
\begin{equation}
\sigma_{\mu\nu} = \frac{1}{2}\bigl[u_{\mu;\alpha} P^\alpha_\nu +
u_{\nu;\alpha} P^\alpha_\mu \bigr] - \frac{1}{3} \vartheta
P_{\mu\nu}, \label{shearcomp}
\end{equation}
where the projection vector $P$:
\begin{equation}
P^2 = P, \quad P_{\mu\nu} = g_{\mu\nu} - u_\mu u_\nu, \quad
P^\mu_\nu = \delta^\mu_\nu - u^\mu u_\nu. \label{proj}
\end{equation}
In comoving system we have $u^\mu = (1,0,0,0)$. In this case one
finds
\begin{equation}
\vartheta = \frac{\dot a_1}{a_1} + \frac{\dot a_2}{a_2} + \frac{\dot
a_3}{a_3} = \frac{\dot V}{V}, \label{expbvi}
\end{equation}
and
\begin{eqnarray}
\sigma_{1}^{1} &=& \frac{1}{3}\Bigl(-2\frac{\dot a_1}{a_1} +
\frac{\dot
a_2}{a_2} + \frac{\dot a_3}{a_3}\Bigr) =  \frac{\dot a_1}{a_1} - \frac{1}{3} \vartheta, \label{sh11}\\
\sigma_{2}^{2} &=& \frac{1}{3}\Bigl(-2\frac{\dot a_2}{a_2} +
\frac{\dot a_3}{a_3} +
\frac{\dot a_1}{a_1}\Bigr) =  \frac{\dot a_2}{a_2} - \frac{1}{3} \vartheta, \label{sh22}\\
\sigma_{3}^{3} &=& \frac{1}{3}\Bigl(-2\frac{\dot a_3}{a_3} +
\frac{\dot a_1}{a_1} + \frac{\dot a_2}{a_2}\Bigr) =  \frac{\dot
a_3}{a_3} - \frac{1}{3} \vartheta. \label{sh33}
\end{eqnarray}
One then finds
\begin{equation}
\sigma^ 2 = \frac{1}{2}\biggl[\sum_{i=1}^3 \biggl(\frac{\dot
a_i}{a_i}\biggr)^2 - \frac{1}{3}\vartheta^2\biggr] =
\frac{1}{2}\biggl[\sum_{i=1}^3 H_i^2 -
\frac{1}{3}\vartheta^2\biggr]. \label{sheargen}
\end{equation}

The Hubble constant of the model is defined by
\begin{equation}
H = \frac{\dot a}{a} = \frac{1}{3} \Bigl(\frac{\dot a_1}{a_1} +
\frac{\dot a_2}{a_2} + \frac{\dot a_3}{a_3}\Bigr) = \frac{1}{3}
\frac{\dot V}{V}. \label{Hubblebv}
\end{equation}
The deceleration parameter $q$, and the average anisotropy parameter
$A_m$ are defined by
\begin{eqnarray}
q &=& - \frac{a \ddot a}{\dot a^2} = 2 - 3\frac{V \ddot V}{\dot V^2} , \label{decparv}\\
A_m &=& \frac{1}{3}\sum_{i=1}^3 \Bigl(\frac{H_i}{H} - 1\Bigr)^2,
\label{anisvi}
\end{eqnarray}
where $H_i$ are the directional Hubble constants:
\begin{equation}
H_1 = \frac{\dot a_1}{a_1}, \quad H_2 = \frac{\dot a_2}{a_2}, \quad
H_3 = \frac{\dot a_3}{a_3}. \label{dirhubc}
\end{equation}

\section{Solution to the field equations}

From \eqref{03bv} immediately follows
\begin{equation}
a_1a_2 = k_1 a_3^2, \quad k_1 = {\rm const.} \label{abcrelbv}
\end{equation}

We also impose use the proportionality condition, widely used in
literature. Demanding that the expansion is proportion to a
component of the shear tensor, namely
\begin{equation}
\vartheta = N_1 \sigma_1^1.\label{propconv}
\end{equation}
The motivation behind assuming this condition is explained with
reference to  Thorne \cite{thorne67}, the observations of the
velocity-red-shift relation for extragalactic sources suggest that
Hubble expansion of the universe is isotropic today within $\approx
30$ per cent \cite{kans66,ks66}. To put more precisely, red-shift
studies place the limit
\begin{equation}
\frac{\sigma}{H} \leq 0.3, \label{propconviexp}
\end{equation}
on the ratio of shear $\sigma$ to Hubble constant $H$ in the
neighborhood of our Galaxy today. Collins et al. (1980) have pointed
out that for spatially homogeneous metric, the normal congruence to
the homogeneous expansion satisfies that the condition
$\frac{\sigma}{\theta}$ is constant.

On account of \eqref{expbvi} and \eqref{sh33} we find
\begin{equation}
a_1 = N_0 V^{ \frac{1}{3} + \frac{1}{N_1}}, \quad N_0 = {\rm const.}
\label{a1v}
\end{equation}

In view of \eqref{Vdef} and \eqref{a3v} from \eqref{abcrelbv}  we
find
\begin{eqnarray}
a_2 &=& \frac{k_1^{1/3}}{ N_0} V^{\frac{1}{3} - \frac{1}{3N_1}}, \label{a2v}\\
a_3 &=& \frac{1}{k_1^{1/3}}  V^{\frac{1}{3}}. \label{a3v}
\end{eqnarray}

Thus, we have derived metric functions in terms of $V$. In order to
find the equation for $V$ we take the following steps. Subtractions
of \eqref{11bv} from \eqref{22bv}, \eqref{33bv} from \eqref{33bv},
and \eqref{33bv} from \eqref{11bv} on account of \eqref{a1v},
\eqref{a2v} and \eqref{a3v} give
\begin{subequations}
\label{V123v}
\begin{eqnarray}
\frac{\ddot V}{V} &=& \frac{\kappa N_1}{2}
 [T_2^2 - T_1^1], \label{V12v}\\
\frac{\ddot V}{V}  &=&
-\kappa N_1[T_3^3 - T_2^2], \label{V23v}\\
\frac{\ddot V}{V}  &=&
-\kappa N_1 [T_1^1 - T_3^3]. \label{V31v}\\
\end{eqnarray}
\end{subequations}
From \eqref{V123v} immediately follows
\begin{equation}
\frac{1}{2}[T_2^2 - T_1^1] = =[T_3^3 - T_2^2] = =[T_1^1 - T_3^3].
\label{T123v}
\end{equation}
After a little manipulation, it could be established that
\begin{equation}
T_1^1 + T_2^2 = 2T_3^3 . \label{T123vn}
\end{equation}
Hence, the energy momentum tensor can be taken as

\begin{eqnarray}
T_\alpha^\beta &=& {\rm diag}[\varepsilon,\,-p_x,\,-p_y,\,-p_z],\nonumber \\
&=& {\rm diag}[1,\,-\omega_x,\,-\omega_y,\,-\omega_z]\varepsilon,\nonumber \\
&=& {\rm diag}[1,\,-(\omega + \delta),\,-(\omega -
\delta),\,-\omega]\varepsilon. \label{emtbvi}
\end{eqnarray}

Thus we conclude that under the proportionality condition, the
energy-momentum distribution of the model should obey \eqref{T123v}.

As one sees, in order to find $V$ we have to impose some additional
condition. Let us apply the law of variation for Hubble parameter
given by \cite{Berman} which yields a constant value of deceleration
parameter. Here, the law reads as
\begin{equation}
H = D a^{-n} = D V^{-n/3}, \label{berman}
\end{equation}
where $D > 0$ and $ n \geq 0$ are constants. Such type of relations
have firstly been considered by \cite{Berman,Gomide} for solving FRW
models. Latter on many authors have used this law to study FRW and
Bianchi type models. In view of \eqref{Hubblebv} and \eqref{berman}
we find
\begin{equation}
\frac{\dot V}{V} = 3D V^{-n/3} \label{Vbv}
\end{equation}
with the solution
\begin{equation}
V = (nD t + C_1)^{3/n}, \quad n \ne 0, \quad C_1 = {\rm const.}
\label{Vbvn}
\end{equation}

\begin{figure}[ht]
\centering
\includegraphics[height=70mm]{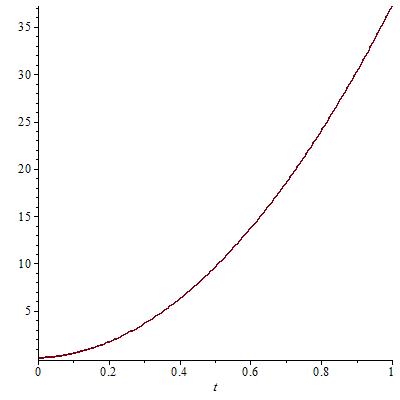} \\
\caption{Evolution of the Universe given by a BV cosmological model.
} \label{V-BV}.
\end{figure}

Fig. [\ref{V-BV}] shows the evolution of the Universe. As one sees,
it is an expanding one.

The value of deceleration parameter is found to be
\begin{equation}
q = n - 1, \label{decparbv}
\end{equation}
which is a constant. The sign of $q$ indicates whether the model
inflates or not. The positive sign of $q$ i.e. $(n > 1)$ correspond
to ``standard" decelerating model whereas the negative sign of $q$
i.e. $0 \leq n < 1$ indicates inflation. It is remarkable to mention
here that though the current observations of SNe Ia and CMBR favours
accelerating models ($ q < 0$), but both do not altogether rule out
the decelerating ones which are also consistent with these
observations \cite{Vishwakarma}.

\section{Physical aspects of Dark energy model}

Let us now find the expressions for physical quantities.

Inserting \eqref{Vbv} into \eqref{Hubblebv} one finds the expression
for expansion $\vartheta$, Hubble parameter $H$:
\begin{equation}
\vartheta = 3H = \frac{3D}{nD t + C_1}, \label{Hubblev1}
\end{equation}

\begin{figure}[ht]
\centering
\includegraphics[height=70mm]{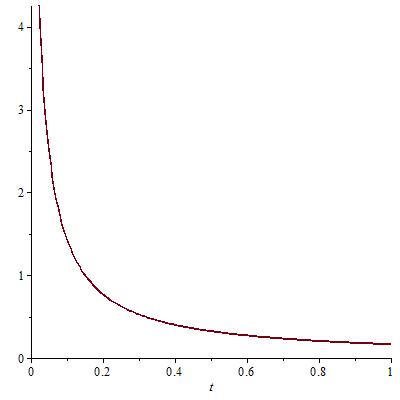} \\
\caption{Evolution of the Hubble parameter. } \label{H-BV}.
\end{figure}

Fig. [\ref{H-BV}] shows the evolution of the Hubble parameter. As
one sees, it is a decreasing function of time.

The value of deceleration parameter is found to be
\begin{equation}
q = n - 1, \label{dpbv}
\end{equation}
which is a constant. The sign of $q$ indicates whether the model
inflates or not. The positive sign of $q$ i.e. $(n > 1)$ correspond
to ``standard" decelerating model whereas the negative sign of $q$
i.e. $0 \leq n < 1$ indicates inflation. It is remarkable to mention
here that though the current observations of SNe Ia and CMBR favours
accelerating models ($ q < 0$), but both do not altogether rule out
the decelerating ones which are also consistent with these
observations \cite{Vishwakarma}.

The anisotropy parameter $A_m$ has the expression
\begin{equation}
A_m = \frac{6}{N_1^2}. \label{aniso}
\end{equation}

The directional Hubble parameters are
\begin{equation}
H_1 = \Bigl(\frac{1}{3} + \frac{1}{N_1}\bigr)\frac{3D}{nDt+C_1},
\quad H_2 = \Bigl(\frac{1}{3}
-\frac{1}{N_1}\bigr)\frac{3D}{nDt+C_1}, \quad H_3 =
\frac{D}{nDt+C_1}. \label{dirhubv}
\end{equation}

From \eqref{00bv} we find the expression for energy density For
energy density in this case we have
\begin{equation}
\varepsilon = \frac{X_1}{(nDt+C_1)^2} - \frac{3 m^2
C_1^2}{(nDt+C_1)^{2/n}}, \label{vebv}
\end{equation}
where $X_1 = 9D^2(1/3 - 1/N_1^2)$. The EoS parameter in this case
has the form
\begin{equation}
\omega =\frac{X_2/(nDt+C_1)^2 +  m^2 C_1^2/(nDt+C_1)^{2/n}
}{X_1/(nDt+C_1)^2 - 3 m^2 C_1^2/(nDt+C_1)^{2/n}}, \label{EoSbv}
\end{equation}
where $X_2 = X_1 - 2D^2(3 - n)$.

\begin{figure}[ht]
\centering
\includegraphics[height=70mm]{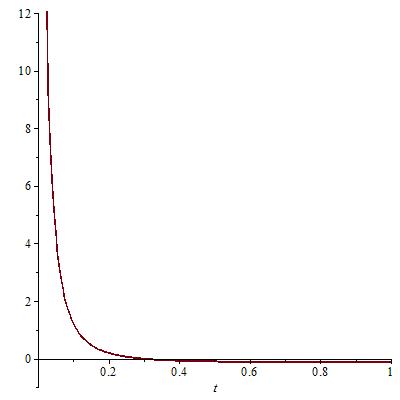} \\
\caption{Evolution of energy density. } \label{Epsilon-BV}.
\end{figure}

Fig. [\ref{Epsilon-BV}] shows the evolution of energy density. As
one sees, it is a decreasing function of time and beginning some
moment of time it may be negative as well.

\begin{figure}[ht]
\centering
\includegraphics[height=70mm]{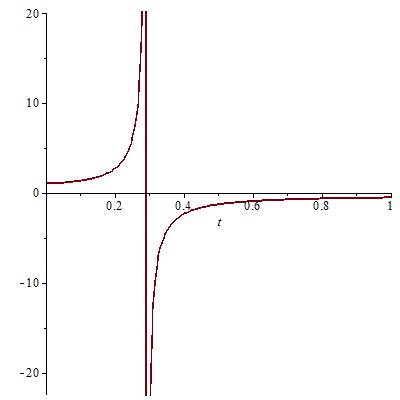} \\
\caption{Evolution of the EoS parameter. } \label{Omega-BV}.
\end{figure}

Fig. [\ref{Omega-BV}] shows the evolution of the EoS parameter. As
one sees, it is a time varying function and changes its sign in the
course of evolution.

From equation \eqref{EoSbv}, it is observed that the equation of
state parameter $\omega$ is time dependent, it can be function of
redshift $z$ or scale factor $a$ as well. The redshift dependence of
$\omega$ can be linear like
\begin{equation}
 \omega(z) =  \omega_{0} + \omega^{'} z,\label{redsh}
\end{equation}
 with $\omega^{'}$ = $\frac{d\omega}{dz}|_{z=0}$ (see Refs. \cite{Huterer,Weller}
 or nonlinear as \cite{Chevallier,Linder}
\begin{equation}
\omega(z) = \omega_{0} + \frac{\omega_{1}z}{1 + z}. \label{redshnl}
\end{equation}
 So, as for as the scale
factor dependence of $\omega$ is concern, the parametrization
\begin{equation}
\omega(a) = \omega_{0} + \omega_{a}(1 - a), \label{redshpar}
\end{equation}
where $\omega_{0}$ is the present value ($a = 1$) and $\omega_{a}$
is the measure of the time variation $\omega^{'}$ is widely used in
the literature \cite{Linder1}.

Let us now compare the our results with the experimental results
obtained in \cite{Knop,Tegmark1,Hinshaw,Komatsu}. It enable us to
conclude that the limit of $\omega$ provided by equation
\eqref{EoSbv} may accommodated with the acceptable range of EoS
parameter. Also it is observed that at $t = t_{c}$, $\omega$
vanishes, where $t_{c}$ is a critical time given by
\begin{equation}
 t_{c} = \frac{1}{nD}\biggl[ \biggl(\frac{X_2
}{m^2C_1^2}\biggr)^{n/2(n - 1)} - C_1\biggr].\label{tcbv}
\end{equation}
Thus, for this particular time, our model represents a dusty
universe. We also note that the earlier real matter at $t \leq
t_{c}$, where $\omega \geq 0$ later on at $t > t_{c}$, where $\omega
< 0$ converted to the dark energy dominated phase of universe.

For the value of $\omega$ to be in consistent with observation
\cite{Knop}, we have the following general condition
\begin{equation}
 t_{1} < t < t_{2}, \label{t12v}
\end{equation}
where
\begin{equation}
 t_{1} = \frac{1}{nD}\biggl[ \biggl(\frac{
X_2 + 1.67 X_1}{-4.01m^2C_1^2}\biggr)^{n/2(n-1)} -
C_1\biggr].\label{t1v}
\end{equation}
and
\begin{equation}
 t_{1} = \frac{1}{nD}\biggl[ \biggl(\frac{
X_2 + 0.62 X_1}{-0.86m^2C_1^2}\biggr)^{n/2(n-1)} -
C_1\biggr].\label{t2v}
\end{equation}

For this constrain, we obtain  $-1.67 < \omega < -0.62$, which is in
good agreement with the limit obtained from observational results
coming from SNe Ia data \cite{Knop}.

For the value of $\omega$ to be in consistent with observation
\cite{Tegmark1}, we have the following general condition

\begin{equation}
t_{3} < t < t_{4}, \label{t34v} \end{equation} where

\begin{equation}
 t_{1} = \frac{1}{nD}\biggl[ \biggl(\frac{
X_2 + 1.33 X_1}{-2.99m^2C_1^2}\biggr)^{n/2(n-1)} -
C_1\biggr].\label{t3v}
\end{equation}

and
\begin{equation}
 t_{1} = \frac{1}{nD}\biggl[ \biggl(\frac{
X_2 + 0.79 X_1}{-1.37m^2C_1^2}\biggr)^{n/2(n-1)} -
C_1\biggr].\label{t4v}
\end{equation}

For this constrain, we obtain  $-1.33 < \omega < -0.79$, which is in
good agreement with the limit obtained from observational results
coming from SNe Ia data \cite{Tegmark1}.

For the value of $\omega$ to be in consistent with observation
\cite{Hinshaw,Komatsu}, we have the following general condition
\begin{equation}
t_{5} < t < t_{6},\label{t56v}
\end{equation}
where
\begin{equation}
 t_{1} = \frac{1}{nD}\biggl[ \biggl(\frac{
X_2 + 1.44 X_1}{-3.32m^2C_1^2}\biggr)^{n/2(n-1)} -
C_1\biggr].\label{t5v}
\end{equation}

and
\begin{equation}
 t_{1} = \frac{1}{nD}\biggl[ \biggl(\frac{
X_2 + 0.92 X_1}{-1.76m^2C_1^2}\biggr)^{n/2(n-1)} -
C_1\biggr].\label{t6v}
\end{equation}

For this constrain, we obtain  $-1.44 < \omega < -0.92$, which is in
good agreement with the limit obtained from observational results
coming from SNe Ia data \cite{Hinshaw,Komatsu}.

We also observed that if

\begin{equation}
 t_{1} = \frac{1}{nD}\biggl[ \biggl(\frac{
X_2 +  X_1}{-2 m^2C_1^2}\biggr)^{n/2(n-1)} - C_1\biggr].\label{t0v}
\end{equation}

then for $t = t_0$ we have $\omega = -1$, i.e., we have universe
with cosmological constant. If $t < t_0$ the we have $\omega > -1$
that corresponds to quintessence, while for $t > t_0$ we have
$\omega > -1$, i.e., Universe with phantom matter \cite{Caldwell1}.

From \eqref{vebv} we found that the energy density is a decreasing
function of time and $\varepsilon \ge 0$ when

\begin{equation}
t \ge \frac{1}{nD}\biggl[ \biggl(-\frac{ X_1
}{m^2C_1^2}\biggr)^{n/2(n-1)} - C_1\biggr]. \label{deposv}
\end{equation}

In absence of any curvature, matter energy density $\Omega_m$ and
dark energy density $\Omega_\Lambda$ are related by the equation

\begin{equation}
\Omega_m + \Omega_\Lambda = \frac{\varepsilon}{3 H^2} +
\frac{\Lambda}{3 H^2} = 1. \label{OmegamL}
\end{equation}

Inserting \eqref{Hubblev1} and \eqref{vebv} into \eqref{OmegamL} we
find the cosmological constant as
\begin{equation}
\Lambda =  \frac{3D^2 - X_1}{(nD t + C_1)^2}  + \frac{3 m^2
C_1^2}{(nDt+C_1)^{2/n}}, \label{Lambdav}
\end{equation}
As we see, the cosmological function is a decreasing function of
time and it is always positive when

\begin{equation}
t \ge \frac{1}{nD}\biggl[ \biggl(\frac{X_1 - 3 D^2}{3 m^2
C_1^2}\biggr)^{n/2(n - 1)} - C_1\biggr]. \label{Lamposv}
\end{equation}

\begin{figure}[ht]
\centering
\includegraphics[height=70mm]{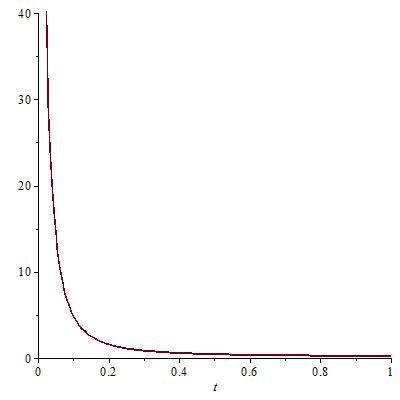} \\
\caption{Evolution of the EoS parameter. } \label{Lambda-BV}.
\end{figure}

Fig. [\ref{Lambda-BV}] shows the evolution of the cosmological
constant. As one sees, it is a time varying function and decreases
with time.

Recent cosmological observations  suggest the existence of a
positive cosmological constant $\Lambda$ with the magnitude
$\Lambda(G\hbar/c^{3})\approx 10^{-123}$. These observations on
magnitude and red-shift of type Ia supernova suggest that our
universe may be an accelerating one with induced cosmological
density through the cosmological $\Lambda$-term. Thus, the nature of
$\Lambda$ in our derived DE model is supported by recent
observations.

For the stability of corresponding solutions, we should check that
our models are physically acceptable. For this, the velocity of
sound is less than that of light, i.e.,

\begin{equation}
0 \le v_s = \frac{dp}{d\varepsilon} < 1. \label{accon}
\end{equation}

\begin{figure}[ht]
\centering
\includegraphics[height=70mm]{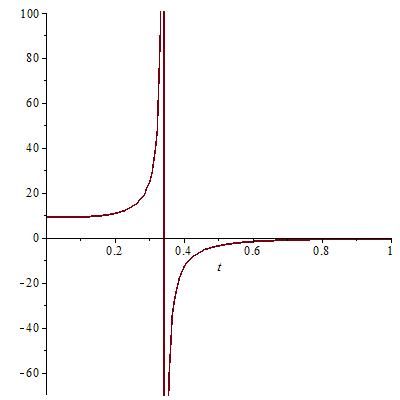} \\
\caption{Speed of sound with respect to cosmic time} \label{Vs-BV}.
\end{figure}

In this case we find

\begin{equation}
v_s = \frac{dp}{d\varepsilon}  = \frac{nX_2 +   m^2 C_1^2
(nDt+C_1)^{2 -2/n} }{nX_1 - 3  m^2 C_1^2(nDt+C_1)^{2 - 2/n}}.
\label{vsv}
\end{equation}
Fig. [\ref{Vs-BV}] shows the behavior of $v_s$ in time.

As one sees, there are regions, where the solution is stable. Fig.
[\ref{Vs-BV}] shows that the solution becomes unstable during the
transition from deceleration to acceleration phase of evolution.
Choosing the problem parameters, such as $n,D$ we can obtain the
stable solutions before or after the transition.

\section{Conclusion}
In this report we have studied the evolution of the universe filled
with dark energy within the scope of a Bianchi type-V model. Exact
solutions to the field equations are obtained using the
proportionality condition and variational law of Hubble parameter.
It was found that the assumption of diagonal energy-momentum tensor
together with the non-diagonal Einstein equation leads to some
restriction on the energy momentum tensor, namely, $T_1^1 + T_2^2 =
2T_3^3$. The behavior of EoS parameter $\omega$ is thoroughly
studied. It is found that the solution becomes stable as the
Universe expands.

\vskip 2mm

\noindent {\bf Acknowledgments}\\
This work is supported in part by a joint Romanian-LIT, JINR, Dubna
Research Project, theme no. 05-6-1060-2005/2013.

\end{document}